\documentclass[numberedappendix,apj]{emulateapj}

\usepackage{amsmath}
\slugcomment{Accepted}
\shortauthors{Bernet et. al.}

\begin{document}
\title{The interpretation of Rotation Measures in the presence of inhomogeneous foreground screens}

\author{M. L. Bernet,  F. Miniati, S.J. Lilly}
\affil{Physics Department, ETH Zurich, Wolfgang-Pauli-Strasse 27, CH-8093 Zurich, Switzerland}
\email{mbernet, fm, simon.lilly@phys.ethz.ch}

\begin{abstract}
  We analyze the redshift evolution of the Rotation Measure (RM) in
  \cite{Taylor2009} dataset, which is based on NVSS radio data at 21
  cm, and compare with results from our previous
  work~\citep{Kronberg2008,Bernet2008,Bernet2010}, based on RMs
  determined at lower wavelengths, e.g. 6 cm.  We find that, in spite
  of the same analysis, Taylor's dataset produces neither an increase
  of the RM dispersion with redshift as found in~\citet{Kronberg2008},
  nor the correlation of RM strength with MgII absorption lines found
  in~\cite{Bernet2008}.  We develop a simple model to understand the
  discrepancy. The model assumes that the Faraday Rotators, namely the
  QSO's host galaxy and the intervening MgII host galaxies along the
  line of sight, contain partially inhomogeneous RM screens. We find
  that this leads to an increasing depolarization towards longer
  wavelengths and to wavelength dependent RM values.  In particular,
  due to cosmological redshift, observations at fixed wavelength of
  sources at different redshift are affected differently by
  depolarization and are sensitive to different Faraday active
  components. For example, at 21 cm the polarized signal is averaged
  out by inhomogeneous Faraday screens and the measured RM mostly
  reflects the Milky Way contributions for low redshift QSOs, while
  polarization is relatively unaffected for high redshift
  QSOs. Similar effects are produced by intervening galaxies acting as
  inhomogeneous screens.  Finally, we assess the performance of
  Rotation Measure synthesis on our synthetic models and conclude that
  the study of magnetic fields in galaxies as a function of cosmic
  time will benefit considerably from the application of such a
  technique, provided enough instrumental bandwidth. For this purpose,
  high frequency channels appear preferable but not strictly
  necessary.

\end{abstract}


\section{Introduction} \label{introduction:sec}

Faraday Rotation Measures (RM) is one of the very few
 methods to probe extragalactic magnetic fields.  
The RM is given by the 
change in observed polarization angle, $\Delta \chi_0$, over a change in 
the observed wavelength square, $\Delta\lambda _0^{2}$.
For a  polarized radio source at cosmological redshift $z_{s}$ 
it is defined as

\begin{equation}
\label{eq:defRM}
RM(z_{s}) = \frac{{\Delta \chi_0 }}{{\Delta \lambda _0^{2} }} = 8.1
\cdot
10^{5} \int\limits_{z_{s}}^{0} \frac{n_e
(z)B_\parallel(z)}{(1+z)^{2}}\frac{dl}{dz}dz,
\end{equation}

where the RM is in units of rad m$^{-2}$, the free electron number
density, $n_{e}$, is in cm$^{-3}$, the magnetic field component 
along the line of sight, $B_{\|}$, is in Gauss,
and the comoving path increment per unit redshift, $dl$/$dz$, is in pc.
Eq. \ref{eq:defRM} assumes a uniform
RM screen across the source and a spatial separation of the
linearly polarized source and the Faraday rotating plasma.

In \cite{Kronberg2008} (K08) we used a sample of 268 RM values of
extragalactic radio sources to assess the redshift evolution in the RM
dispersion. We found an increase in the RM dispersion with redshift,
which became statistically significant above $z \sim 1$. We postulated
that the increase in the RM dispersion is produced by magnetic fields
in intervening galaxies. To test this hypothesis spectra of 71 QSOs at
UVES/VLT were taken. In \cite{Bernet2008} (B08) we showed that indeed
sightlines with intervening strong MgII absorption systems have
significantly higher RM values than those without. The findings in
that work implies that $\sim 10 \; \mu\rm{G}$ magnetic fields exist in
galaxies out to $z \sim 1.3$.

Recently \cite{Taylor2009} (TSS09) determined RM values of 37'543
sources, based on polarization observation of the NRAO VLA Sky Survey
(NVSS). After the release of the RM catalogue we compared our RM
values with those of TSS09 and found major differences between the two
datasets. As we show in section \ref{discrepancies} using the RM
values of TSS09 the results of K08 and B08 can not be reproduced. In
this work we present an analysis of the differences found between the
two datasets. We then develop a toy model to show that such
differences can be accounted for by inhomogeneities in the RM screens
of the QSOs host galaxies and intervening galaxies. In view of this
model, due to the strong depolarization effects, RM data based on low
frequencies observations and interpreted according to
Eq.~(\ref{eq:defRM}) are inadequate to probe inhomogeneous RM screens
produced by intervening galaxies.

Recently there has been an increased attention to the effects of
inhomogeneous RM screens on the observed degree of polarization and
polarization angle as function of
wavelength. \cite{Rossetti2008,Mantovani2009} used multiwavelength
radio observations to model depolarization of their sources. They
showed that the depolarization as a function of wavelength can be
better described including a covering factor for the inhomogeneous
screens. \cite{Farnsworth2011} used Westerbork Synthesis Radio
Telescope (WSRT) observation at $\sim 1 \rm{m}$ combined with NVSS
observation at 21 cm to test different depolarization models and RM
determination methods. They considered the traditional Faraday
dispersion screen \citep{Burn1966}, a two component model for Faraday
Rotation which produces oscillation in the degree of polarization, and
the RM Synthesis method. The comparison of the different RM values
obtained from these methods revealed that if Faraday structure is
present, the different methods may lead to different RMs. They further
stress the importance of considering both the polarization angle and
amplitude for the correct determination of the RMs.

While several effects may be at work, as briefly discussed in section
\ref{effects:sec}, in this paper we keep the level of sophistication
of the model to a minimum and focus on the role of Faraday screen
particularly on simple ideas associated with inhomogeneity which can
account for important observational effects and test these ideas for
consistency with available data.  The rest of this paper is organized
as follows.  In section \ref{RM_data} the RM datasets of B08 and K08
are compared and the differences are presented. Previous work on
depolarization of extragalactic sources is shortly presented in
section \ref{sec:Depmodel}. In section \ref{modelling} a
depolarization toy model is presented which can account for the
differences in the RM values of the two datasets.  A discussion of the
nature of the partial inhomogeneous RM screens and computed Faraday
spectra as comparison for RM surveys is presented in section
\ref{sec:discussion}. In section \ref{sec:summary} we give a summary of our findings.

\section{Rotation Measure Data}
\label{RM_data}
\subsection{Datasets}
\label{datasets}

The RM values used in K08, B08 and \cite{Bernet2010} were collected by P. Kronberg
and collaborators over the past decades using polarization observations at various telescopes, including VLA and Effelsberg. The sample consists of 901 sources with determined RM and redshifts, partly taken by Kronberg and collaborators, partly taken from the literature 
\citep{SimardNormandin1981}.
Until the work of TSS09 it has been the only existing large RM catalogue.
Generally at least three wavelengths were used  for the RM determination (Phil Kronberg, private communication) using polarization data between 37 to 0.9 GHz, with the bulk of the data between $\sim$ 10.5 to 1.6 GHz \citep{SimardNormandin1981}. 

In the following sections we will assume that the determinations of the RM values of K08 were typically done at 6 cm and designate the RM values as $\rm{RM_{6}}$. Of course due to the heterogeneity of the sample the wavelength range and typical value over which the RMs are determined might vary.
While this is undesirable, its main effect should be the introduction of noise in the relations 
predicted by our model between measured quantities.
There might be additional issues associated with the heterogenous character of the polarization 
data used for the RM determinations, which, however, are not addressed here.

The sample of K08 consists of 268 sources at Galactic latitudes $|b| > 45^{\circ}$  (exact definition given in K08).  In B08 we obtained high resolution spectra of 71 relatively bright QSOs and 
relaxed the selection to $|b| > 30^{\circ}$. 
Here we use the subset that was employed in the study of B08 and \cite{Bernet2010} 
(the RM data are still proprietary and will be published elsewhere by P. Kronberg),
except when we look for differences in the redshift evolution and use all sources at $|b| > 30^{\circ}$ which were also in the \cite{Taylor2009} catalogue.

The RM values of TSS09 are based on the NRAO VLA Sky Survey (NVSS)
from \cite{Condon1998} which covers the sky at declinations $|\delta|
> 40^{\circ}$ in Stokes I, Q and U. The survey imaged the sky at 21 cm
with a resolution of $\sim 45$ arcsecs and produced a catalogue of
$2\times10^{6}$ discrete sources. TSS09 choose a subsample of this
catalogue and derived RM values of these sources based on
determination of the polarization angles at 1364.9 MHz and 1435.1
MHz. The subsample was selected by requesting a source intensity
$\rm{I} > 5\: \rm{mJy}$ and a 8$\sigma$ detection in polarized
intensity. To ensure that the polarized intensity is not dominated by
instrumental effects they only considered sources with a fractional
polarization greater than 0.5 $\%$, which yielded a RM catalogue of
37'543 objects. In the following sections we will designate the RM
values of TSS09 as $\rm{RM_{21}}$.

\subsection{Discrepancies}
\label{discrepancies}
\begin{figure} 
\begin{center}
\includegraphics[width=0.450\textwidth]{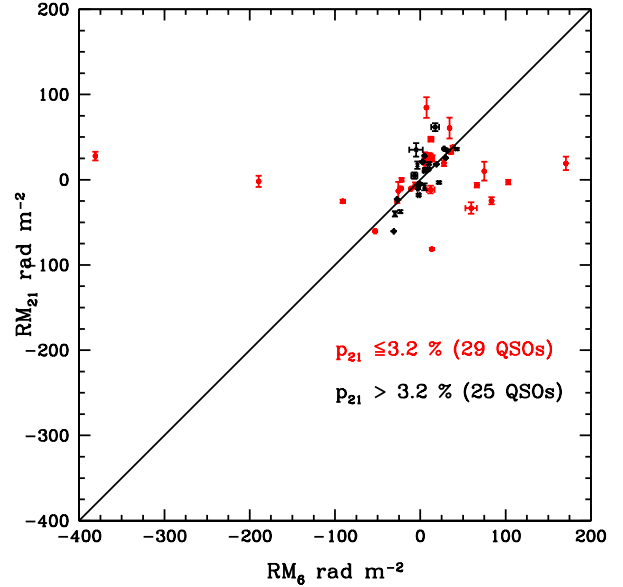}
\caption[Comparison of the RM values]{Comparison of the RM values of K08 and TSS09, $\rm{RM_{6}}$ and $\rm{RM_{21}}$ respectively, splitted according to whether the degree of polarization is above (black crosses) or below (red filled circles) $p_{21}=3.2\%$. The two datasets differ especially for low $p_{21}$ values. Errors on $\rm{RM}_{6}$ are available for 24 of 54 RM measurments. \label{fig:CompPPK_T}}
\end{center}
\end{figure}

\begin{figure*}
\begin{center}
\includegraphics[width=1.0\textwidth]{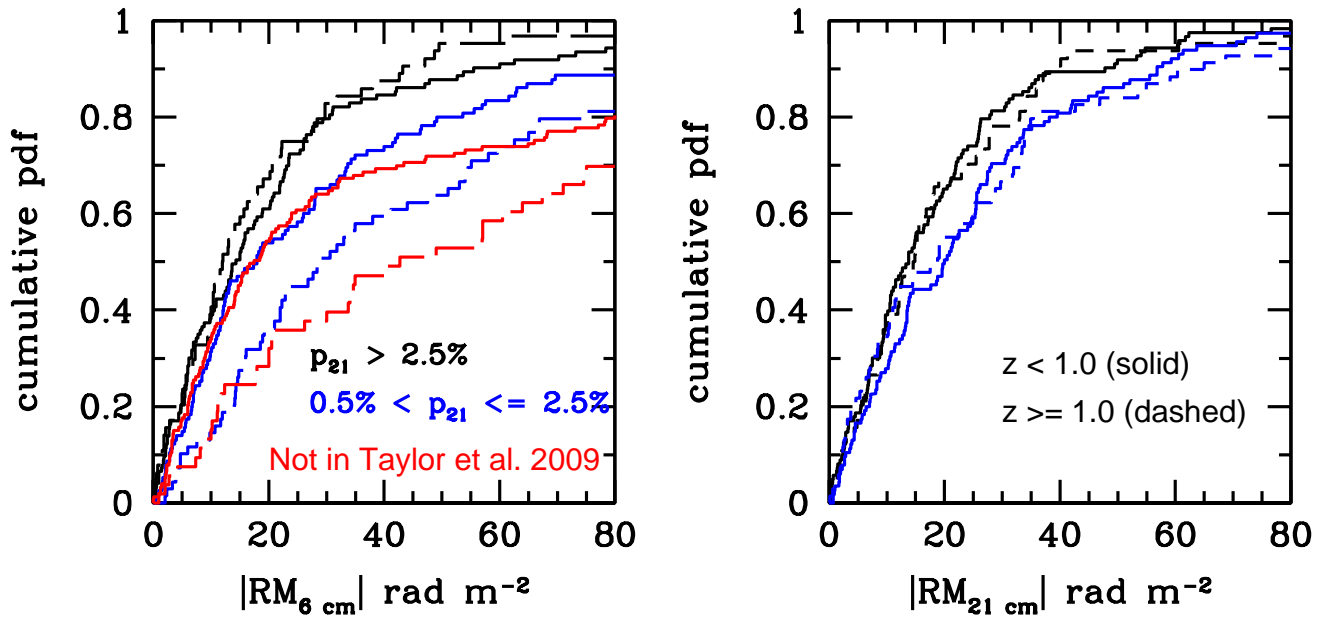}
\caption[  ]{Comparison of the cumulative distributions of RM values measured typically at 6cm (left panel) and at 21 cm (right panel). The samples are split according to the polarized fraction $p_{21}$ and the redshifts of the sources. Significant redshift evolution is seen in the $|\rm{RM_{6}}|$ values but not for the $|\rm{RM_{21}}|$ values. For sources with $0.5\% < p_{21} < 2.5\%$ the RM distributions 
for sources above z=1 is broader than for sources below z=1, at a significance level of 99.75\%. Red histograms are for NVSS sources not included in~\cite{Taylor2009} catalogue. This is because they are below the inclusion limits of the catalogue, which are a 8$\sigma$ detection in polarized intensity and $p_{21} > 0.5\%$. For these sources also a significant redshift evolution can be seen with a significance level of 99.64\%. \label{fig:cumRMz}}
\end{center}
\end{figure*}

\begin{figure}
\begin{center}
\includegraphics[width=0.450\textwidth]{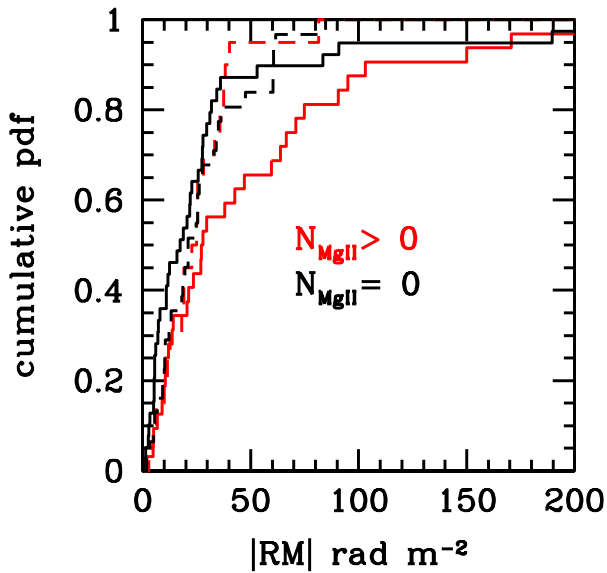}
\caption[Comparison of the RM distribution]{Comparison of the cumulative RM distributions of B08 (solid) and TSS09 (dash) for QSO sight lines with no MgII absorption systems (black lines) or with one or two, respectively (red lines). RM data in TSS09 were available for 54 out of the 71 QSOs in B08.  \label{fig:CumRMdist}}
\end{center}
\end{figure}

After the release of the TSS09's RM catalogue we checked if it contained any of the sources in B08, and found this to be the case for 54 out of 71 of them. 
Somewhat surprisingly we found striking differences in the RM values which are briefly 
presented here:

i) A comparison of the RM values from K08 and TSS09 reveals that the biggest differences in the RM datasets correspond to large RM values of K08 and sources having a low degree of polarization $p_{21}$. Here $p_{21}$ is the degree of polarization at ~ 21 cm from TSS09 (average degree of polarization at the two wavelengths 20.89 cm $\&$ 21.96 cm). The situation is illustrated in Figure \ref{fig:CompPPK_T}. 
Sources with $p_{21} > 3.2\%$ (black cross, 25 QSOs) have quite similar RM values in both datasets and mostly crowd around the diagonal line. On the other hand sources with $p_{21} < 3.2\%$ show a big scatter around the diagonal line. Furthermore there are sources which have large RM values ( $|\rm{RM}| > 50$ rad m$^{-2}$) in the dataset of K08 but very low $|\rm{RM}|$ values in the sample of TSS09. The value of $p_{21} = 3.2\%$ maximizes these effects in Figure \ref{fig:CompPPK_T}.

ii) Using a sample of 371 RM values from TSS09 
at latitudes $|b| > 30^{\circ}$ and for which the QSO's redshifts, $z$, were available, 
we found no increase in the RM dispersion with $z$, contrary to the main finding of K08. The situation is illustrated in Figure \ref{fig:cumRMz}, where the cumulative RM distributions at 6 cm 
(left panel) and 21 cm (right panel) are shown, split at the median fractional polarization of the total sample, $p_{21}=2.5\%$ (blue and black lines), and according to whether the source redshift is below or above $z=1.0$ (solid and dash line, respectively). The cumulative distribution of $|\rm{RM_{21}}|$ 
shows no evolution with $z$ in both low and high $p_{21}$ samples, whereas 
there is a significant broadening of the low $p_{21}$ ($0.5\% < p_{21} < 2.5\%$)
$|\rm{RM_{6}}|$ distribution with $z$. A Kolmogorov-Smirnov test (KS test) reveals that the RM distributions above and below $z=1.0$ are different at a significance level of 99.75\%. We also show  (red distributions) sources with determined $\rm{RM_{6}}$ values which are not in the \cite{Taylor2009} catalogue, but 
have declination $|\delta| > 40^{\circ}$ and are thus in the NVSS. Most likely these sources are below the inclusion limits of the \cite{Taylor2009} catalogue, which are a $8\sigma$ detection in polarized intensity and $p_{21}>0.5\%$. A redshift evolution can be seen also for these sources with a similar significance level of 99.64\%. Most likely the broadening of the RM distribution towards lower polarization $p_{21}$ is due to Galactic depolarization and possibly partly due to an increased error in RM, $\delta_{RM}$, towards lower polarized flux levels. This is supported by the fact that $\delta_{RM} \propto \delta_{P}/P$, where P is the polarized flux and $\delta_{P}$ its error \citep{Brentjens2005}, and the high correlation between $p_{21}$ and polarized flux, as attested by a Kendall's $\tau$ test, which yields $\tau=0.40$ and a chance probability of only $4.4\times10^{-30}$ using all 371 RM values. 
However the broadening with z seen in the $|\rm{RM_{6}}|$ distributions is hard to explain by an observational bias. The interpretation of a $p_{21}$ dependent redshift evolution of the $|\rm{\rm{RM_{6}}}|$ distribution is however not straightforward either. In section \ref{modelling} we show that it is crucial to distinguish between sight lines with/without intervening galaxies.

iii) The result of B08 that the RM distribution for QSOs with strong MgII systems along their sight lines  is broader (at $\sim 95\%$ significance level) than for QSOs free of absorbers, completely disappears using the RM values of TSS09. Figure \ref{fig:CumRMdist} shows the cumulative distribution 
(dash histograms) of $|\rm{RM_{21}}|$ values from TSS09 for sight lines with 
($N_{\rm{MgII}}>0$, black)) and without MgII ($N_{\rm{MgII}} =0$, red)
absorption systems.
Clearly there is no difference between the two RM distributions. For comparison the 
solid histograms show the distributions of the $|\rm{RM_{6}}|$ values of B08.

\subsection{Caveats}\label{effects:sec}

In the following sections we present a simple model to describe the impact of
inhomogeneous, redshift dependent Faraday screens, on the polarization 
and RM of distant QSOs, as a function of wavelength.

However, additional effects may be at work when comparing quantities
measured at different frequencies.  In particular besides potential
large uncertainties measured with both datasets, the low and high
frequency emissions may originate from separate components in the
source (e.g. the radio lobes and the compact core in a radio galaxy),
inducing all sorts of frequency dependent effects, e.g., location,
angular size, spectral indices and fractional polarization.  For
example, in some radio sources the fractional polarization appear to
increase and then decrease with wavelength~\citep{Conway1974}.

Also, while our model could be validated (or ruled out) by comparison
of its predictions with observational data, this would require a
high quality dataset, where the effects discussed above (and possibly
others) are kept under strict control. Unfortunately such a dataset is
not available to us at the moment, which forces us to refrain from
such comparison. The latter, however, might be possible in the near
future, with the delivery of new radio data from polarization surveys,
either planned or just under way (see Sec.~\ref{rms:sec}).

\section{Depolarization by inhomogeneous Faraday screens}

\label{sec:Depmodel}

\subsection{Previous work}
In this section we develop a simple model to understand the effects arising from inhomogeneous Faraday screens and show that these can produce important
depolarization effects that reproduce the differences of the RM datasets of 
K08 and TSS09.

The observed polarization $\mathbf{p}$ can be written as a complex number as:
\begin{equation}
\label{eq:ComplexP}
\mathbf{p}=pe^{2i\chi},
\end{equation}
where $p=P/I$ is the fractional degree of polarization given by the
ratio of the polarized and total intensity, P and I, respectively and $\chi$ is the polarization angle determined by the other Stokes parameters, U and Q, as $\chi=1/2\arctan(U/Q)$.

The linear dependence between the polarization angle, $\chi$ and $\lambda^{2}$ stated in Eq. \ref{eq:defRM} is only valid for the case of a uniform foreground screen. Both the presence of unresolved inhomogeneities in the Faraday screen
and/or sources of polarized radiation embedded within the Faraday active region produce increasing depolarization of the source at longer wavelength \citep{Burn1966,Tribble1991,Sokoloff1998}.
This leads to nonlinear dependencies of the polarization angles and $\lambda^{2}$. In such cases detailed modeling is necessary for the correct interpretation of the RMs and for their use in measurements of magnetic fields.

There are several mechanisms that can reduce the fractional degree of polarization either in the radio source itself or in its foreground:
\begin{itemize}
\item[i)] \cite{Burn1966,Sokoloff1998} showed that random fluctuations in the magnetic field within the source lead to a wavelength independent reduction of the degree of polarization. The latter is given by the ratio of the regular-to-total magnetic field energy, $p_{0}=p_{t}(\gamma)B_{r}^{2}/B^{2}$, where $p_{t}(\gamma)=\frac{3\gamma+3}{3\gamma+7}$ is the theoretical value of the degree of polarization which depends on the spectral index $\gamma$ of the emitting relativistic electrons ($\sim 0.74$ for $\gamma \sim 2.8$), and $p_{0}$ is the degree of polarization as
$\lambda \to 0$.

\item[ii)] A further effect which reduces the degree of polarization is {\it differential Faraday Rotation} \citep{Sokoloff1998}. This effect happens if the synchrotron emission and the medium producing the Faraday rotation are not spatially separated. In such a case polarization angles from different depths within the source are rotated differently. The line of sight integrated emission suffers increasing depolarization with increasing wavelength. For extragalactic sources this effect seems not to be important \citep{Tribble1991}.
 
\item[iii)] Inhomogeneous Faraday Rotation screens within the radio beam lead to depolarization of the sources. If the RM screen is modelled by many independent RM cells, this effect is called depolarization by {\it external Faraday dispersion} \citep{Burn1966,Sokoloff1998}. If the RM screen varies systematically within the radio beam this effect is called {\it beam depolarization}. 

\end{itemize}

\cite{Burn1966} gives the formula
\begin{equation}
\label{eq:BurnDep}
p=p_{0}\exp(-2\sigma_{RM}^{2}\lambda^{4})
\end{equation}
to describe the depolarization induced by an inhomogeneous Faraday
screen with a RM dispersion $\sigma_{RM}$. 
Assuming for simplicity that each RM cell is a cube of linear size $l_{0}$,
then each cell contributes a dispersion $\sigma_{c}=0.81Bn_{e}l_{0}$
and $\sigma_{RM}=\sigma_{c}\sqrt{\eta}$,
where $\eta=L/l_{0}$ is the number of cells along the
path-length L traversed by the radio waves.

The size of the cell is determined by the scale above which the RM
contributions are uncorrelated and can be
determined by computing the structure function in high resolution RM
maps.\footnote{The structure function D at scale s is
  defined as $D(s)=\langle(RM(x+s)-RM(x))^{2}\rangle$ and
  $\langle\rangle$ means ensemble averaging.}

\cite{Rossetti2008} and \cite{Mantovani2009} modelled polarisation
observations of $\sim 65$ compact steep spectrum sources between 2.8
to 21 cm done with the WSRT, VLA and Effelsberg telescope. They
observed that the fractional degree of polarisation at large
wavelengths is too large to be explained by Burn's depolarization law
(Eq.\ref{eq:BurnDep}). They observed that for a large fraction of
sources p remains approximately constant above 6 - 13 cm. To account
for these observation they suggested that just a fraction of the
polarised source is covered by a depolarising inhomogeneous RM screen
and modified Eq.~\ref{eq:BurnDep} to

\begin{equation}
\label{eq:fracDep}
p=p_{0}(f_{c}\exp(-2\sigma_{RM}^{2}\lambda^{4}) +(1-f_{c})),
\end{equation}
where $f_{c}$ is the covering factor of the source. 

Here we emphasize that \textit{partial} coverage of the polarized
source by inhomogeneous RM screens is the key for explaining the
differences in the RM values of K08 and TSS09.

\section{Modelling}
\label{modelling}
\subsection{Extension to cosmological screens}

In Eq. \ref{eq:fracDep} the ($1+z)^{-2}$ correction for cosmological distances is not included. 
Assuming that the RM dispersion causing the depolarization, $\sigma_{RM}$, is constant with z, 
we modify Eq. \ref{eq:fracDep} for cosmological sources as:
\begin{equation}
\label{eq:fracDep_z}
p=p_{0}(f_{c}\exp(-2\sigma_{RM}^{2}(1+z)^{-4}\lambda^{4}) +(1-f_{c})).
\end{equation}
Thus the width of $p(\lambda)$ changes as a function of z as

 \begin{equation}
\label{eq:widthp}
\sigma_{p}=\frac{(1+z)^{2}}{2\sigma_{RM}}.
\end{equation} 
This shows that depolarization by a non-evolving rest frame Faraday
dispersion screen is expected to decrease for screens at higher
redshift. Below, we explicitly compute the depolarization by typical
RM screens as a function of redshift of the sources.

\subsection{A simple model}

In our model the Faraday screens consist of three components:
(i) an inhomogeneous foreground screen local to the source with covering 
factor $f_{c,QSO}$, (ii) an inhomogeneous foreground
screen in intervening galaxies with covering factors $f_{c,MgII}$ and 
(iii) a homogeneous screen due to the Milky Way, which is assumed
uniform across the extension on the sky of the polarized source.

The uniform Faraday screen in the Milky Way is set to a constant 
value, $RM_{MW}= -10\: \rm{rad\: m^{2}}$. This value is consistent with \cite{Schnitzeler2010} who determined the
Milky Way contribution of Rotation Measures at $|b| > 20^{\circ}$ using the \cite{Taylor2009} data.

On the other hand, each inhomogeneous foreground Faraday screen is
characterized by an ensemble of cells of size $l_{0}$ with
uncorrelated RM values.  The RM values have a Gaussian distribution
with dispersion $\sigma_{RM_x}/(1+z_x)^{2}$ and zero mean, where
$\sigma_x$ is the rest-frame RM dispersion, $z_x$ the screen's redshift
and, $x=QSO,MgII,$ labels different screens types, for which the 
above parameters may differ.

 If the RM screen is at the source redshift and the source has linear
 size  $s$, then it is covered by $N=s^{2}/l_{0}^{2}$ cells.  Of these, only
 a fraction $f_{c,x}$ will be Faraday active for each screen. If the RM screen
is at the redshift of the MgII systems, $s$ is the projected linear size of the source 
viewed from the Earth at the distance to the MgII system.

Each screen is then fully described by a realization of $N$ RM values,
a fraction $f_{c,x}$ of which are extracted from the associated
Gaussian distribution to characterize the active cells, while the
reminder are set to zero to represent the inactive cells. Assuming a
uniform flux of normalized intensity (I=1) and uniform (zero)
intrinsic polarization angle across the source surface, we can write
the following relations for the remaining non-zero Stokes parameter U,
Q, the angle and degree of polarization, respectively,
\begin{align}
U(\lambda^{2}) &= \frac{1}{N}\sum_{i=1}^{N}\sin\left(2RM_{i}\lambda^{2}\right) \\
Q(\lambda^{2}) &= \frac{1}{N}\sum_{i=1}^{N}\cos\left(2RM_{i}\lambda^{2}\right) \\
\label{eq:polangle}
\chi(\lambda^{2}) & = \frac{1}{2}\arctan\left(\frac{U}{Q}\right)  \\ 
\label{eq:degreepol}
p(\lambda^{2})&= \sqrt{U^{2}+Q^{2}},
\end{align} 
where
\begin{equation}
\label{eq:RMi}
RM_{i}=RM_{QSO,i}+RM_{MgII,i}+RM_{MW},
\end{equation} 
are random variables determined by a Monte-Carlo realization.

\begin{figure} 
\begin{center} 
\includegraphics[width=1\hsize]{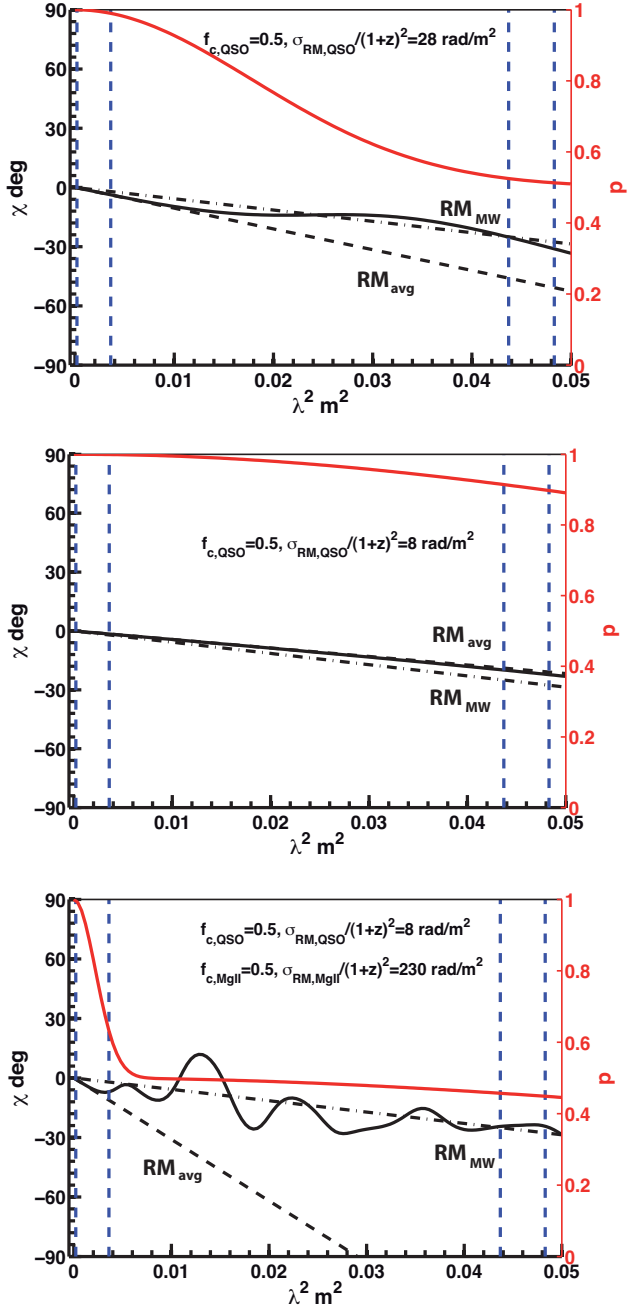}
\caption[Summary of effects of intrinsic RM screen]{
Angle of polarization, $\chi$ (black solid line -- left y-axis),
and theoretical polarization fraction, p (red solid line -- right y-axis), as a function of $\lambda^{2}$,
for three models of inhomogeneous Faraday screens:
a low (top) and high (mid) redshift QSO with no intervening absorbing system,
and a high redshift QSO with intervening absorber (bottom). 
   The dash line indicates the rotation of 
   polarization angles due to the average RM within the beam,
  $\chi=RM_{avg}\lambda^{2}$, and the dash-dot line 
  that due to the Milky Way, $\chi=RM_{MW}\lambda^{2}$.
  The vertical blue dash lines from left to
  right, indicate typical wavelength of observations at 1.5 cm, at 6 cm (K08), 
  and at 20.89 cm and 21.96 cm (NVSS survey, \cite{Condon1998})
  for \cite{Taylor2009} calculated RM values. 
  In the top and mid panel the Faraday screens include an inhomogeneous
  screen local to the source with a covering factor $f_{c}=0.5$
  and a screen in the Milky Way that is uniform across the source extension.  In the top
  panel the source QSOs is at $z < 1.0$ and has 
  $\sigma_{RM,QSO}/(1+z)^{2}= 28\; \rm{rad\: m^{-2}}$; in the mid panel the
  QSO is at $z > 1.0$ and has $\sigma_{RM,QSO}/(1+z)^{2} =
  8\; \rm{rad\: m^{-2}}$. The bottom panel is the same as the mid panel with an additional 
  Faraday screen due to an intervening system, with $\sigma_{RM,MgII}/(1+z)^{2}\sim\: 230\:
  \rm{rad\: m^{-2}}$ and $f_{c}=0.5$. \label{fig:sketch_dep}}
\end{center}
\end{figure}

In Figure \ref{fig:sketch_dep} we present generic results for three
representative cases, namely a low ($z_{QSO}<1$) and high
($z_{QSO}>1$) redshift QSO with no intervening absorber in the top and
mid panel, respectively, and a high redshift QSO with intervening
absorber in the bottom panel. More detailes are specified below.  In
all cases, the inhomogeneous screen comprise eight cells,
$N\,f_{c,x}=8$.

In each panel we plot the polarization angle, $\chi$ (black solid line
-- left y-axis), and the polarization fraction, p (red solid line --
right y-axis) as a function of $\lambda^{2}$.
To emphasize the importance of the effects of depolarization, we also
show the rotation in polarization angle produced by (a) the average RM
within the beam (dash line),
$$\chi=RM_{avg}\lambda^{2},$$
where
\begin{align} \label{avgrm:eq} 
RM_{avg}=&\frac{1}{N}\sum_{i=1}^{N}RM_{i}\\  \label{avgrm2:eq}
=&RM_{MW}+\frac{1}{N}\sum_{i=1}^{N}RM_{QSO,i}+RM_{MgII,i},
\end{align}
and (b) by the Milky Way (dash-dot line), 
$$\chi=RM_{MW}\lambda^{2}.$$
Note that $RM_{avg}$ is in general non-zero and different
from $RM_{MW}$. This is because for fixed fraction of active cells 
the second term in Eq.~\ref{avgrm2:eq}
is a random number of aplitude $1/\sqrt{N}$, with null
probability of being either zero or $-RM_{MW}$.
Finally, from left to right, the vertical blue dash lines indicate the typical 
observational wavelength of 20 GHz for \citep{Jackson2010,Murphy2010}, of 5 GHz for K08 and of 1.435 GHz and 1.365 GHz
for the NVSS survey \citep{Condon1998}) and~\cite{Taylor2009}. 

The top panel illustrates the case of a low redshift sources ($z < 1.0$),
with observed RM dispersion of $\sigma_{RM,QSO}/(1+z)^{2}= 28\;
\rm{rad\: m^{-2}}$, $f_{c,QSO}=0.5$ and no intervening MgII absorber ($f_{c,MgII}=0$). 
It can be seen that at
short wavelengths, $\lambda^{2} < 0.014$, application of the simple 
$\lambda^2$ relation in Eq. \ref{eq:defRM}, yields the average RM within the beam
from which the RM distributions in the foreground screens can be inferred \citep{Bernet2008}. 
However, at longer wavelengths the flux through the inhomogeneous screen is depolarized so that the 
change in polarization angle is dominated by the 
contribution of the Milky Way. Correspondingly,
the degree of polarization is given by $p \sim 1-f_{c,QSO}$.

The mid panel illustrates the case for a high redshift source, $z > 1.0$
with observed RM dispersion of $\sigma_{RM,QSO}/(1+z)^{2}= 8\;
\rm{rad\: m^{-2}}$ and $f_{c,QSO}=0.5$. In this case, depolarization is 
negligible up to $\lambda^{2}=0.05$ or at 1.4 GHz. This means across
most of the considered wavelength range application of the  
$\lambda^2$ relation in Eq. \ref{eq:defRM} would yield 
the average RM contributed by both the QSO screen and the Milky Way.

Finally, the bottom panel illustrates the same case as the mid panel but
with an additional Faraday screen due to an absorber, with
$\sigma_{RM,MgII}/(1+z)^{2}\sim\: 230\: \rm{rad\: m^{-2}}$ and $f_{c,MgII}=0.5$.

This case has direct connection to the study of B08,
who showed that intervening galaxies traced by MgII absorption
lines contribute an additional Faraday screen with an observed rest
frame RM dispersion $\sigma_{MgII}= 140\;\rm{rad\; m^{-2}}$. 
In the context of our simple model,
this implies a RM dispersion of the individual cells of the screen, 
responsible for the depolarization,
$\sigma_{RM,MgII}=\sigma_{MgII}\sqrt{N/f_{c,MgII}}$.
The value for $\sigma_{RM,MgII}/(1+z)^{2}$ assumed above is 
is based on this relation and the choice of $N\,f_{c,MgII}=8$.
The bottom panel of Fig.~\ref{fig:sketch_dep} 
shows that with this choice of parameters the intervening screen causes 
significant depolarization above $\lambda^2 \sim 0.003 \rm{m^{2}}$.
In this case the application of the $\lambda^2$ relation at longer wavelengths  
yields the Milky Way RM value.

The approach taken here, is valid in the limit of large number of random cells.
Using the total flux - angular diameter relation given by
\cite{Windhorst1984} we can estimate the extension of the region of the polarized emission. 
The median flux of our sources which are in the \cite{Taylor2009} catalogue (371 sources)
is 1.2 Jy. This translates to a median source size in total flux of $\sim$ 10 arcsecs. Since the
polarized emission is likely to be dominated by some hotspots the effective size in
polarized emission is expected to be smaller than that.
Given that the extension of the polarized emission is a few arcsecs 
and what we know about the scale of fluctuations of the RM (see next section) we 
expect a significant number of cells within the beam. 
In addition, using Eq. \ref{eq:fracDep_z} and future multi-wavelength radio polarization
observations one can determine $\sigma_{RM,MgII}$ and $f_{c,MgII}$ directly
and thus calculate N. 

Finally, we have assumed a homogeneous RM screen in the Milky Way.
\cite{Gaensler2005}, however, shows that arcsec scale fluctuation
exist in the RM screen of the Large Magellanic Cloud
and~\cite{Haverkorn2008} shows that there are RM fluctuation in the
local interstellar medium.  Inhomogeneous Galactic RM contribution,
which might vary for different sight lines, would lead to additional
depolarization and add scatter to the predicted trends (see section
\ref{Dep_model}) of our simple model, but cannot mimic them.

\subsection{Model predictions} \label{Dep_model}

Based on this toy model with the free parameters $f_{c,x}$,
$\sigma_{RM,x}$, we can predict simple trends in the observed angle
and degree of polarization measured at fixed frequency as a function
of radio source's redshift.  For simplicity we shall assume there is
no redshift evolution in the model parameters.  We must also
distinguish between the two cases with and without the Faraday screen
provided by intervening galaxies.

For the case without intervening galaxies the model predicts
\begin{itemize}

\item[i)] that towards low redshifts $p_{21}/p_{\lambda \rightarrow 0}
  \sim 1-f_{c,QSO}$, because the flux through the inhomogeneous Faraday
  screen is depolarized.

\item[ii)] that towards high redshift $p_{21}/p_{\lambda \rightarrow 0} \approx  1 $ 
because the depolarization effects are suppressed. However, 
$p_{21}/p_{\lambda \rightarrow 0}$ anti-correlates with $\sigma_{RM,QSO}$ due to residual
depolarization effects associated with the local screen.

\item[iii)] that towards low redshifts the discrepancy $|\rm{RM_{\lambda \rightarrow 0}-RM_{21}}|$ increases 
because due to depolarization effects 
$\rm{RM_{\lambda \rightarrow 0}}$ measures $\rm{RM_{avg}}$ while $\rm{RM_{21}}$ is dominated 
by the Milky Way contribution. 

\item[iv)] that the discrepancy $|\rm{RM_{\lambda \rightarrow
      0}-RM_{21}}|$ decreases towards high redshifts because both
  $\rm{RM_{\lambda \rightarrow 0}}$ and $\rm{RM_{21}}$ tend to measure
  the average $\rm{RM_{avg}}$ (Eq.~\ref{avgrm:eq}).

\end{itemize}
For the case with intervening galaxies the model predicts
\begin{itemize}

\item[i)] that at all redshift the discrepancy $|\rm{RM_{\lambda \rightarrow 0}-RM_{21}}|$ is large, 
because due to depolarization effects 
$\rm{RM_{\lambda \rightarrow 0}}$ measures $\rm{RM_{avg}}$ and $\rm{RM_{21}}$ is dominated 
by the Milky Way contribution.

\item[ii)] that at all redshift $p_{21}/p_{\lambda \rightarrow
    0}\lesssim 1-f_{c,MgII}$ because the sources will be significantly
  depolarized.

\end{itemize}

\section{Discussion}
 \label{sec:discussion}

\subsection{Nature of the partial inhomogeneous RM screen}

It is instructive to look at high resolution Very Long Baseline Array
(VLBA) radio polarization observations of some of our objects in
K08. Examples of sources where VLBA polarization observations exist
are 3C43 \citep{Cotton2003}, 3C118 \citep{Mantovani2009} and B1524-026
\citep{Mantovani2002}. Typical resolutions of these observations are
$\sim 8 \times 8\: \rm{mas^{2}}$, which corresponds to a resolution of
$64\:\rm{pc}\times 64\:\rm{pc}$ at $z \sim 1$. At such a high
resolution complex structures in polarization angles and RM maps are
revealed.  Often it can be seen, e.g B1524-026 \citep{Mantovani2002}
that there is a dominant compact component in polarised flux and a
more extended diffuse polarised component. Further it can be seen that
the diffuse component consist of many independent RM cells. That means
that for unresolved observations at large wavelengths the diffuse
component will cancel and the more compact component dominates the
observations.

For some sources it possible that the number of cells
$f_{c}N$ in the RM screen is very low. In this case no depolarization
is observed but p and $\chi$ oscillate. This situation was observed 
for the source 3C 27 by \cite{Goldstein1984}.  \cite{Rossetti2008} fitted a
two component model to the data to describe p and $\chi$ for the
source B3 0110+401. \cite{Farnsworth2011} also used a two component
model to describe radio polarization observation at large wavelengths
$\lambda \sim 1$ m.

For the sightlines with intervening galaxies it is very natural to
assume that the magnetic fields within them will lead to
depolarization, an effect studied in the LMC by \cite{Gaensler2005}.
 Using the Milky Way as typical galaxy we would expect
depolarization of the background QSOs. For NGC 1310 this effect was observed by 
\cite{Fomalont1989} and~\cite{Schulman1992}.
One possibility is that the
source covers both the spiral arms and interarm regions. For the
spiral arms the coherence lengths of the magnetic fields are much
shorter as suggested by observations of \cite{Haverkorn2008}. These
authors gives an outer scale of turbulence of 0.1 kpc for interarm
regions and 10 pc for the spiral arms in the Milky Way\footnote{The
  small coherence lengths of the magnetic fields has probably to do
  with our point of view in the disc of the Milky Way. We know of
  course from observations of nearby spiral galaxies that there is a
  magnetic field component which is coherent over several kpc.}. In
this way the area covered by the spiral arms would be depolarized and
the interarms would produce a coherent screen. Other interpretations
are again as for the QSO itself that the polarized emission is
dominated by one compact component which competes with a more extended
diffuse emission. If the size of the compact component is small
enough, this component could get a coherent screen whereas the
extended source gets different RM screens which will lead to
depolarization of this component.
 
\subsection{Rotation Measure Synthesis} \label{rms:sec}

At the moment there are several polarization surveys under way or
planned, e.g. GALFACTS \citep{Taylor2010}, LOFAR \citep{Anderson2012}, 
or POSSUM \citep{Gaensler2010} within ASKAP. These
polarization surveys should pave the way for the planned polarization
survey for SKA \citep{Gaensler2004,Gaensler2009,Beck2004b}. In
particular one wishes to study the evolution of magnetic fields in
galaxies and in the intergalactic medium as a function of cosmic time
using large RM datasets. 

\begin{figure} 
\begin{center}
\includegraphics[width=0.48\textwidth]{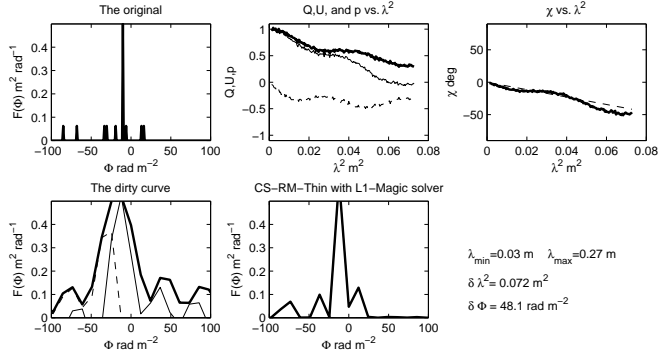}
\caption[FD_spectrum_lowzQSO]{Original Faraday spectrum for a low
  redshift QSO and its associated observed polarization angle and
  degree, together with the recovered Faraday spectrum for a finite
  wavelength coverage. The Monte Carlo realization is identical to the
  top panel in Figure \ref{fig:sketch_dep}. Upper left panel: Original
  Faraday spectrum for a low redshift QSO with a homogeneous Milky Way
  RM screen of $\rm{RM}=-10\: \rm{rad\: m^{2}}$ and an inhomogeneous
  intrinsic RM screen with $\sigma_{RM,QSO}/(1+z)^{2}= 28\: \rm{rad\:
    m^{-2}}$ and a covering factor $f_{c}=0.5$. Upper middle panel:
  Corresponding U and Q Stokes parameters (thin solid line, thin
  dashed line) and degree of polarization p (thick solid line) as a
  function of $\lambda^{2}$. An observational error
  $\sigma_{U,Q}\approx 0.017\%$ is added to every channel. Upper right
  panel: Corresponding polarization angle $\chi$
  vs. $\lambda^{2}$. The dashed line indicates the expected
  polarization angle from the homogeneous Milky Way RM
  contribution. Lower left panel: Recovered Faraday spectrum for the
  wavelength range $\lambda_{min}=0.03 \rm{m}$ to $\lambda_{max}=0.27
  \rm{m}$. The thin solid/dashed line indicates the real/imaginary
  part of the spectrum and the thick solid line shows the
  amplitude. Lower middle panel: Faraday spectrum after applying the
  CS-RM-Thin method of \cite{Li2011} to correct for the limited
  wavelength coverage.  \label{fig:FD_spec_lowzQSO}}
\end{center}
\end{figure}

\begin{figure} 
\begin{center}
\includegraphics[width=0.50\textwidth]{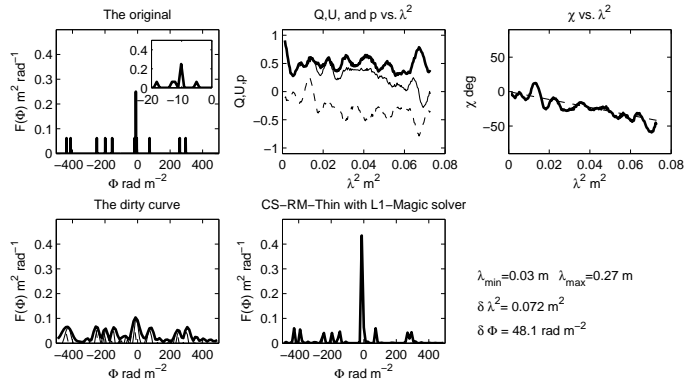}
\caption[FD_spectrum_highzQSO_MgII]{Identical to Figure \ref{fig:FD_spec_lowzQSO} but for the case of a high redshift QSO with an intervening MgII system. The Monte Carlo realization is identical to the bottom panel in Figure \ref{fig:sketch_dep}\label{fig:FD_spec_highzQSO_MgII}}
\end{center}
\end{figure}

\begin{figure} 
\begin{center}
\includegraphics[width=0.49\textwidth]{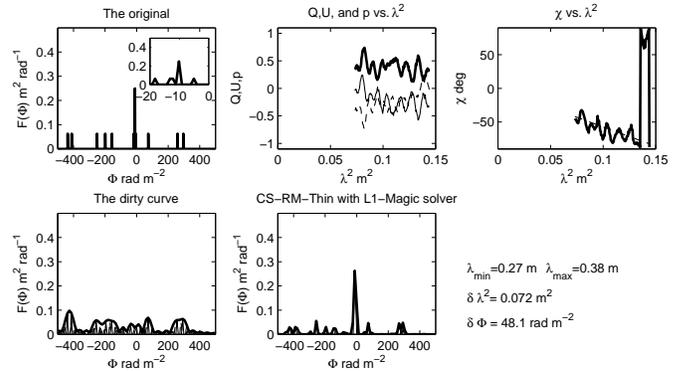}
\caption[FD_spectrum_highzQSO_MgII_largew]{Identical to Figure \ref{fig:FD_spec_highzQSO_MgII} with same $\Delta \lambda^{2}$ coverage and thus identical Faraday depth resolution $\delta \phi$ but with observations done at longer wavelengths. \label{fig:FD_spec_highzQSO_MgII_largew}}
\end{center}
\end{figure}

With the large number of available spectral channels in these surveys,
e.g. a few thousands, Faraday Rotation Measure Synthesis
\citep{Brentjens2005} can be performed. Using this technique one
performs a Fourier transformation of the observed complex polarization
$p(\lambda^{2})$ to obtain a Faraday depth spectrum $F(\phi)$. Here
the case of a single uniform Faraday screen corresponds to a delta
function in Faraday depth space and in this case the Faraday depth is
equal to the traditional RM (Eq.~\ref{eq:defRM}).

Basic parameters for a Faraday survey are the maximum observable
Faraday depth, $\phi_{max} \approx \sqrt{3}/ \delta\lambda^{2}$, the
Faraday depth resolution, $\delta \phi \approx 2\sqrt{3}/ \Delta
\lambda^{2}$ and the largest scale in Faraday depth that one is
sensitive to, max-scale $\approx \pi/\lambda_{min}^{2}$. Here
$\delta\lambda^{2}$ is the channel width, $\Delta \lambda^{2}$ is the
total bandwidth and $\lambda_{min}^{2}$ is the smallest wavelength
(squared) of the observations. Regions which produce only Faraday
rotation but do not emit polarized radiation can be described by Dirac
$\delta$ functions in $\phi$ space and are Faraday-thin sources
\citep{Brentjens2005}. Below we discuss our model of the
depolarization of QSOs in terms of Faraday Rotation Measure
Synthesis. This is to provide some basic comparisons for ongoing or
future RM surveys. See e.g. \cite{OSullivan2012} for current observations 
of Faraday depth spectra. 

In the next, with the auxilium of
Fig.~(\ref{fig:FD_spec_lowzQSO})--(\ref{fig:FD_spec_highzQSO_MgII_largew}) we
give a simple example of the application of RM-synthesis to the same
cases studied in Figure~\ref{fig:sketch_dep}, which illustrates the
advantages of this emerging technique.  In Figure
\ref{fig:FD_spec_lowzQSO} the original Faraday spectrum for a low
redshift QSO without intervening galaxies is shown. Here the Monte
Carlo realization is the same as in Figure \ref{fig:sketch_dep} (upper
panel) with adopted parameters $\sigma_{RM,QSO}/(1+z)^{-2}= 28\:
\rm{rad\: m^{-2}}$ and $f_{c,QSO}=0.5$. In particular, all of the  16 cells have the
contribution from the MW and eight cells have an additional
intrinsic low-z QSO contribution. The Faraday spectrum is only
real valued which means that the intrinsic polarization angles are
zero. The adopted minimum and maximum wavelengths are
$\lambda_{min}=0.03\: \rm{m}$ and $\lambda_{max}=0.27\; \rm{m}$ and
correspond to a Faraday depth resolution of $\delta \phi \approx\: 48.1\:
\rm{rad\: m^{-2}}$. The choosen wavelength range corresponds to the
offered wavelength range at the Australia Telescope Compact Array
which is similar to the EVLA \citep{Beck2012}.  The recovered Faraday
spectrum with this resolution is shown in the lower left panel. 

There are several methods to recover the information lost due to the
incomplete wavelength coverage \citep{Heald2009,Li2011}. We use here
the compressive sampling (CS) method of \cite{Li2011}. Best results
are obtained if prior information about the Faraday spectrum is
present, e.g. if the sources are Faraday thin or thick. The obtained
Faraday spectrum after applying the CS RM thin method of \cite{Li2011}
to the Q and U vectors (upper middle panel) is shown in the lower
right panel. It can be seen that the homogeneous RM component at -10
$\rm{rad\: m}^{-2}$ can be approximately recovered, but not all of the
inhomogeneous RM components.

In Figure \ref{fig:FD_spec_highzQSO_MgII} the Faraday spectrum for a
high redshift QSO with an intervening galaxy is shown. Here the
Faraday spectrum is identical to the one used in Figure
\ref{fig:FD_spec_lowzQSO} in the lower panel with the parameters
$\sigma_{RM,MgII}/(1+z)^{2}\sim\: 230\: \rm{rad\: m^{-2}}$ and
$f_{c,MgII}=0.5$, $\sigma_{RM,QSO}/(1+z)^{2}\sim\: 8\: \rm{rad\:
  m^{-2}}$ and $f_{c,QSO}=0.5$, and $\rm{RM}=-10\: \rm{rad}\: m^{-2}$.
In the Monte Carlo realization
shown in Figure \ref{fig:FD_spec_highzQSO_MgII} all of the 16 cells have the
contribution from the MW, eight (random) cells have an additional
intrinsic high-z QSO contribution and eight (random) cells have an 
additional contribution from the intervening galaxy. 
With the observational parameters $\lambda_{min}=0.03\: \rm{m}$ and
$\lambda_{max}=0.27\: \rm{m}$ and $\delta \phi \approx 48.1 \; \rm{rad\: m^{-2}}$ 
the MW+QSO components can not be resolved but lead to a
distinct peak at $\phi \approx -10\; \rm{rad}\; \rm{m}^{-2}$ with an
amplitude $\sim 0.45$ $\rm{m}^{2}\: \rm{rad^{-1}}$ in the recovered
Faraday spectrum. On the other hand, the eight RM components with
contributions from intervening galaxies can be well resolved with the
assumed covered wavelength range.

For comparison in Figure \ref{fig:FD_spec_highzQSO_MgII_largew}
we show the recovered Faraday spectrum for observations using longer
wavelengths, $\lambda_{min}=0.27\: \rm{m}$ and $\lambda_{max}=0.38\:
\rm{m}$ but with the same covered wavelength range $\Delta
\lambda^{2}$. This shows that the qualitative features of the
Faraday spectrum are also recovered, although with a lower quality
than in Fig.~\ref{fig:FD_spec_highzQSO_MgII}.
 
In summary, from section \ref{Dep_model} using polarization angle
$\chi$ vs. $\lambda^{2}$ observations one might conclude that it is
best to do observations at short wavelengths (below the exponential
fall off) in order to measure the average RM. However as we have
illustrated above using RM Synthesis this is not necessary. In order
to be able to resolve the individual RM components a large $\Delta
\lambda^{2}$ is required. (See \cite{Beck2012} for a summary of the
Faraday depth resolution $\delta \phi$ of current and future radio
telescopes.) The proposed wavelength range for SKA with $\lambda \sim
0.03 - 4.3\: \rm{m}$ and $\Delta \lambda \sim 18\: \rm{m^{2}}$ and
$\delta \phi \sim\: 0.2\: \rm{rad}\: \rm{m}^{-2}$ will offer a superb
Faraday resolution to resolve individual RM components in intervening
galaxies.

\section{Summary and conclusions}
\label{sec:summary}

In previous work we have used RM values from K08, typically determined
at 6 cm, to study magnetic fields in normal galaxies at high
redshift. When the analysis was repeated using the recently released
RM data of TSS09, determined at 21 cm, we could find neither a
correlation between $|$RM$|$ and absorbing systems as in B08, nor an
increase in the RM dispersion with z as found in K08.  Motivated by
the above results, we have attempted to understand those differences
in terms of a simple model based on inhomogeneous Faraday screens
associated both with the QSO and the MgII host galaxies, and for both
low and high redshift.

We find that the presence of inhomogeneous screens leads to important
departures from the classical $\lambda^2$ dependence of the rotation
of the polarization angle.  Related to this are depolarization effects
which become stronger towards higher wavelengths.  As a result, due to
cosmological redshift, observations at fixed wavelength are affected
differently by depolarization and are sensitive to different Faraday
active components.  In particular depolarization effects become
stronger for lower redshift QSOs.  We find that the depolarization
saturates to a value given by the intrinsic polarization times the
complement of the covering factor of the inhomogeneous screen,
$(1-f_c)$.

Actual predictions depend on the assumed values for the model
parameters. The following results apply for the choices made in
section.~\ref{modelling}, which are relevant for the current
investigation.  When the line of sight to the QSO is free of
absorption systems, application of the $\lambda^2$ analysis to extract
RM values from radio observations then has the following consequences:
\begin{itemize}
\item for low redshift QSOs, $\rm{RM_{21}}$ is dominated by the Milky
  Way contribution, while $\rm{RM_{\lambda \rightarrow 0}}$ measures
  $\rm{RM_{avg}}$ (see Eq. \ref{avgrm:eq}).
\item for high redshift QSOs, in general the RM value reflect $\rm{RM_{avg}}$
\end{itemize}

The presence of absorption systems contributes an additional Faraday
screen which further depolarizes the low frequency radiation.
Therefore,
\begin{itemize}
\item while $\rm{RM_{\lambda \rightarrow 0}}$ is given by
  $\rm{RM_{avg}}$, $\rm{RM_{21}}$ is dominated by the Milky Way
  contribution and the discrepancy $|\rm{RM_{\lambda \rightarrow
      0}-RM_{21}}|$ is large at all redshifts.
\end{itemize}

In conclusion, while the model is admittedly simple, it seems
plausible to consider that the discrepancy between results based on
K08 and TSS08 RM are due to the severe depolarization induced by
inhomogeneous Faraday screen on high wavelength radiation.  This,
however, does not exclude the importance of other effects.

Finally, the study of magnetic fields in galaxies as a function of
cosmic time will benefit considerably from the application of
RM-synthesis, which has the power to disintangle the contribution from
inhomogeneous magneto-active components. For this purpose,
instrumental bandwidth is most important, although higher frequency
channels appear to deliver higher quality. This conclusions may be 
refined with future investigations.

\acknowledgements We are very grateful to an anonymous referee for
several valuable comments which helped improve the manuscript
considerably.  This work was supported by the Swiss National Science
Foundation and has made use of NASA's Astrophysics Data System.
During the refereeing process a related work by \cite{Arshakian2011}
appeared. It makes general predictions for future radio polarization
surveys in qualitative agreement with ours, though it does not study
the case of intervening galaxies. Further the RM catalogue of \cite{Hammond2012} appeared
with determined redshifts of 4003 sources of the \cite{Taylor2009} sample. Similar to our
findings they do not see any redshift evolution of the RM distribution in their large sample.


\vskip 0.5truecm


\begin{thebibliography}{}

\bibitem[Anderson et al.(2012)]{Anderson2012} Anderson, J., Beck, 
R., Bell, M., et al.\ 2012, arXiv:1203.2467 

\bibitem[Arshakian \& Beck(2011)]{Arshakian2011} Arshakian, T.~G., \& Beck, R.\ 2011, \mnras, 418, 2336 

\bibitem[Beck et al.(2012)]{Beck2012} Beck, R., Frick, P., Stepanov, R., \& Sokoloff, D.\ 2012, \aap, 543, A113 


\bibitem[Beck \& Gaensler(2004)]{Beck2004b} Beck, R., \& Gaensler, B.~M.\ 2004, NewAR, 48, 1289 

\bibitem[Bernet et al.(2008)]{Bernet2008} Bernet, M.~L., Miniati, 
F., Lilly, S.~J., Kronberg, P.~P., 
\& Dessauges-Zavadsky, M.\ 2008, \nat, 454, 302 

\bibitem[Bernet et al.(2010)]{Bernet2010} Bernet, M.~L., Miniati, F., \& Lilly, S.~J.\ 2010, \apj, 711, 380 

\bibitem[Brentjens \& de Bruyn(2005)]{Brentjens2005} Brentjens, M.~A., \& de Bruyn, A.~G.\ 2005, \aap, 441, 1217 

\bibitem[Burn(1966)]{Burn1966} Burn, B.~J.\ 1966, \mnras, 133, 
67  

\bibitem[Condon et al.(1998)]{Condon1998} Condon, J.~J., Cotton, 
W.~D., Greisen, E.~W., Yin, Q.~F., Perley, R.~A., Taylor, G.~B., 
\& Broderick, J.~J.\ 1998, \aj, 115, 1693 

\bibitem[Conway et al.(1974)]{Conway1974} Conway, R.~G., Haves, 
P., Kronberg, P.~P., et al.\ 1974, \mnras, 168, 137 

\bibitem[Cotton et 
al.(2003)]{Cotton2003} Cotton, W.~D., Spencer, R.~E., Saikia, D.~J., \& Garrington, S.\ 2003, \aap, 403, 537 

\bibitem[Farnsworth et al.(2011)]{Farnsworth2011} Farnsworth, D., 
Rudnick, L., \& Brown, S.\ 2011, \aj, 141, 191 

\bibitem[Fomalont et al.(1989)]{Fomalont1989} Fomalont, E.~B., 
Ebneter, K.~A., van Breugel, W.~J.~M., 
\& Ekers, R.~D.\ 1989, \apjl, 346, L17 

\bibitem[Gaensler et al.(2005)]{Gaensler2005} Gaensler, B.~M., 
Haverkorn, M., Staveley-Smith, L., et al.\ 2005, Science, 307, 1610 

\bibitem[Gaensler(2009)]{Gaensler2009} Gaensler, B.~M.\ 2009, IAU 
Symposium, 259, 645 

\bibitem[Gaensler et al.(2004)]{Gaensler2004} Gaensler, B.~M., Beck, 
R., \& Feretti, L.\ 2004, AN, 48, 1003 

\bibitem[Gaensler et al.(2010)]{Gaensler2010} Gaensler, B.~M., 
Landecker, T.~L., Taylor, A.~R., 
\& POSSUM Collaboration 2010, Bulletin of the American Astronomical Society, 42, \#470.13 


\bibitem[Goldstein \& Reed(1984)]{Goldstein1984} Goldstein, S.~J., Jr., \& Reed, J.~A.\ 1984, \apj, 283, 540 

\bibitem[Hammond et al.(2012)]{Hammond2012} Hammond, A.~M., 
Robishaw, T., \& Gaensler, B.~M.\ 2012, arXiv:1209.1438 

\bibitem[Haverkorn et al.(2008)]{Haverkorn2008} Haverkorn, M., Brown, 
J.~C., Gaensler, B.~M., \& McClure-Griffiths, N.~M.\ 2008, \apj, 680, 362 

\bibitem[Heald et al.(2009)]{Heald2009} Heald, G., Braun, R., \& Edmonds, R.\ 2009, \aap, 503, 409 

\bibitem[Jackson et al.(2010)]{Jackson2010} Jackson, N., Browne, 
I.~W.~A., Battye, R.~A., Gabuzda, D., 
\& Taylor, A.~C.\ 2010, \mnras, 401, 1388 

\bibitem[Kronberg et al.(2008)]{Kronberg2008} Kronberg, P.~P., 
Bernet, M.~L., Miniati, F., Lilly, S.~J., Short, M.~B., 
\& Higdon, D.~M.\ 2008, \apj, 676, 70 

\bibitem[Kronberg et al.(2012)]{Kronberg2012} Kronberg et al., in preparation 

\bibitem[Li et al.(2011)]{Li2011} Li, F., Brown, S., Cornwell, T.~J., \& de Hoog, F.\ 2011, \aap, 531, A126 

\bibitem[Mantovani et 
al.(2002)]{Mantovani2002} Mantovani, F., Junor, W., Ricci, R., Saikia, D.~J., Salter, C., \& Bondi, M.\ 2002, \aap, 389, 58 

\bibitem[Mantovani et 
al.(2009)]{Mantovani2009} Mantovani, F., Mack, K.-H., Montenegro-Montes, F.~M., Rossetti, A., \& Kraus, A.\ 2009, \aap, 502, 61 

\bibitem[Murphy et al.(2010)]{Murphy2010} Murphy, T., et al.\ 
2010, \mnras, 402, 2403 

\bibitem[O'Sullivan et al.(2012)]{OSullivan2012} O'Sullivan, S.~P., 
Brown, S., Robishaw, T., et al.\ 2012, \mnras, 2504 

\bibitem[Rossetti et 
al.(2008)]{Rossetti2008} Rossetti, A., Dallacasa, D., Fanti, C., Fanti, R., \& Mack, K.-H.\ 2008, \aap, 487, 865 

\bibitem[Schnitzeler(2010)]{Schnitzeler2010} Schnitzeler, D.~H.~F.~M.\ 
2010, \mnras, 409, L99 

\bibitem[Schulman \& Fomalont(1992)]{Schulman1992} Schulman, E., \& Fomalont, E.~B.\ 1992, \aj, 103, 1138 

\bibitem[Simard-Normandin et al.(1981)]{SimardNormandin1981} 
Simard-Normandin, M., Kronberg, P.~P., \& Button, S.\ 1981, \apjs, 45, 97 

\bibitem[Sokoloff et al.(1998)]{Sokoloff1998} Sokoloff, D.~D., 
Bykov, A.~A., Shukurov, A., et al.\ 1998, \mnras, 299, 189 

\bibitem[Taylor et al.(2009)]{Taylor2009} Taylor, A.~R., Stil, 
J.~M., \& Sunstrum, C.\ 2009, \apj, 702, 1230 

\bibitem[Taylor \& Salter(2010)]{Taylor2010} Taylor, A.~R., \& Salter, C.~J.\ 2010, Astronomical Society of the Pacific Conference Series, 438, 402 

\bibitem[Tribble(1991)]{Tribble1991} Tribble, P.~C.\ 1991, \mnras, 
250, 726 


\bibitem[Windhorst et al.(1984)]{Windhorst1984} Windhorst, R.~A., van Heerde, G.~M., \& Katgert, P.\ 1984, \aaps, 58, 1 



\end{thebibliography}
\end{document}